\documentclass[
 reprint,
 amsmath,amssymb,
 aps,
 prl,
]{revtex4-2}

\usepackage{graphicx}% Include figure files
\usepackage{dcolumn}% Align table columns on decimal point
\usepackage{bm}% bold math
\usepackage{comment}
\usepackage{color}

\begin{document}

\preprint{APS/123-QED}

\title{%\underline{\small{\tt ver. 0.5 (\today)}}\vspace{5mm}\\
Negative magnetoresistance and sign change of the planar Hall effect due to the negative off-diagonal effective-mass in Weyl semimetals}

\author{Akiyoshi Yamada}
\author{Yuki Fuseya}%
\affiliation{%
 Department of Engineering Science, University of Electro-Communications, Chofu, Tokyo 182-8585, Japan
}%


%\affiliation{
%\affiliation{
%\affiliation{%


\date{\today}% It is always \today, today,

\begin{abstract}
We theoretically investigated the magnetoresistance (MR) and planar Hall effect (PHE) in Weyl semimetals based on the semiclassical Boltzmann theory, focusing on the fine structure of the band dispersion. We identified that the negative longitudinal MR and sign change in the PHE occur because of the negative off-diagonal effective-mass with no topological effects or chiral anomaly physics. Our results highlight the crucial role of the off-diagonal effective-mass, which can cause anomalous galvanomagnetic effects. We propose that the PHE creates a dip in their temperature dependence, which enables the experimental detection of the Weyl point.
\end{abstract}

\maketitle
%\tableofcontents

The galvanomagnetic effect has played a crucial role in solid state physics for many years. For example, the Hall effect is used to determine charge-carrier densities and the transverse magnetoresistance (TMR) is used to calculate mobilities \cite{Quantum_Kittel,Ziman,Beer1963}. Extremely sensitive magnetic sensors have been fabricated using the planar Hall effect (PHE) \cite{Schuhl1995,Tang2003}. The microscopic origin of the galvanomagnetic effect is the Lorentz force, which is solely a classical effect. Hence, one would anticipate that the theoretical investigation of galvanomangetic effects is straightforward without any difficulty. In reality, however, this is not the case. For example, it is often explained that longitudinal magnetoresistance (LMR), where the electric current and magnetic field are parallel, cannot be realized because the Lorentz force does not exist under this condition. By contrast, many materials are known to exhibit LMR \cite{Allgaier1958,Luthi1959,Olsen1967}. The same phenomenon also occurs under the PHE. 
Although the microscopic origin of the phenomena is clearly understood as the Lorentz force, accurately predicting how the macroscopic galvanomagnetic effect is altered in actual materials remains challenging.

Recently, anomalous galvanomagnetic effects, which cannot be interpreted by the conventional theory, have been observed in various materials \cite{Xiaochun2015,Li2016,Liang2018,Vashist2021,Liang2018,Kumar2018,Wu2018,Li2018,Li2019,Yang2020} and have attracted renewed interest from the viewpoint of the chiral anomaly \cite{Nielsen1983,Son2013,Burkov2015,Nandy2017,Burkov2017} or the Berry curvature \cite{Dai2017}.  In Weyl electron systems, the charge conservation is violated between different chiralities under a magnetic field parallel to the electric field; the so-called chiral anomaly \cite{Nielsen1983}. The consequence of the chiral anomaly is as a negative LMR \cite{Son2013,Burkov2015}. Moreover, a secondary consequence of the chiral anomaly is angular oscillation with a period $\pi$ in the PHE where the magnetic field is rotated in the plane parallel to the current \cite{Nandy2017,Burkov2017}. Several experimental studies have suggested the possibility of detecting chiral anomaly based on the aforementioned negative LMR \cite{Xiaochun2015,Li2016,Liang2018,Vashist2021} and PHE \cite{Liang2018,Kumar2018,Wu2018,Li2018,Li2019,Yang2020}.
These anomalous behaviors cannot be interpreted on the basis of {\it simple} semiclassical theory; thus, they are believed to be evidence of chiral anomaly.

\begin{figure}[tb]
\includegraphics[width=8cm]{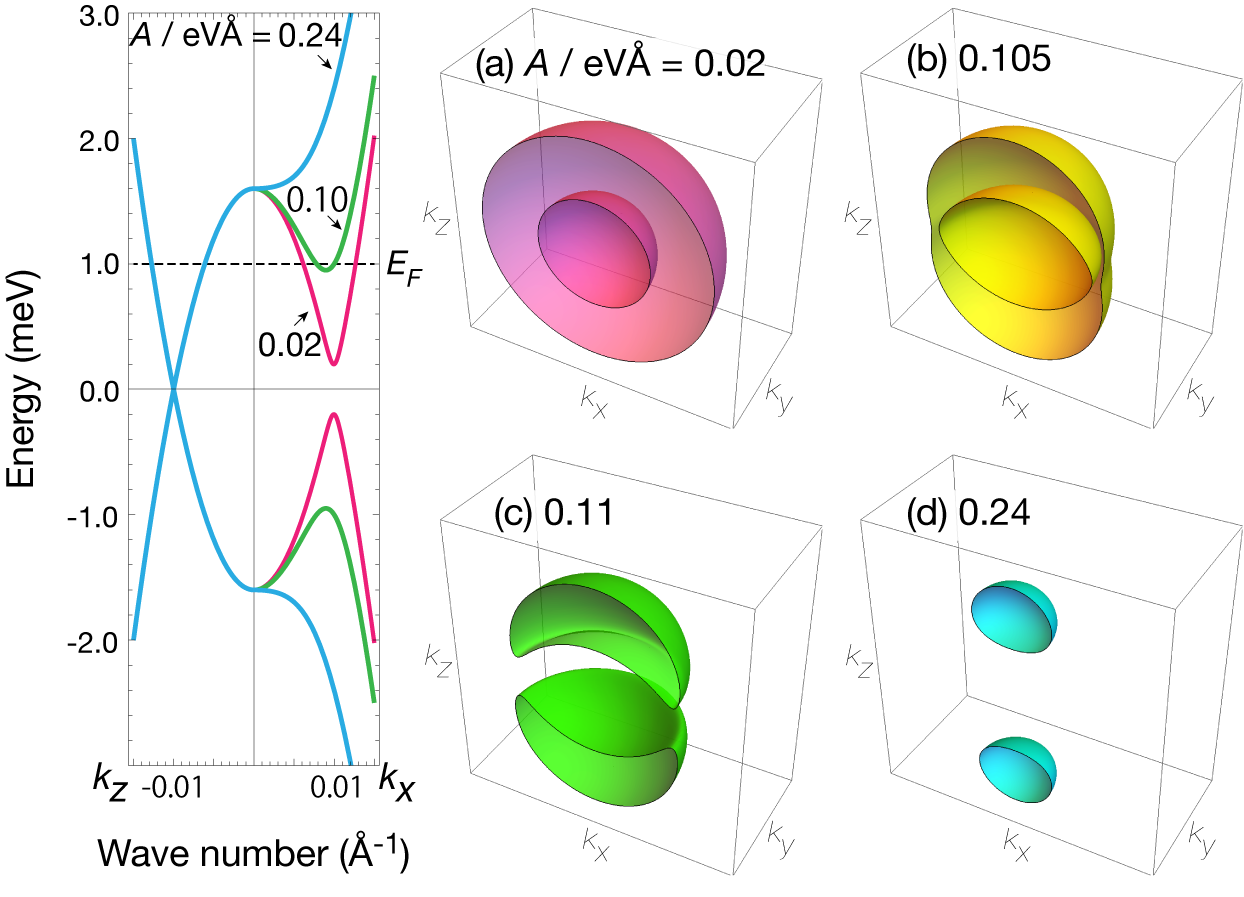}
\caption{\label{fig1} (Left) Band dispersion in the Weyl semimetal model, Eq. \eqref{Hamiltonian} for $A/{\rm eV\AA}=0.02, 0.10, 0.24$. (a)--(d) Cross- sections of the Fermi surface of a Weyl semimetal for $E_F = 1$ meV. (a) semimetal, (b) semimetal near the Lifshitz transition $A=A_0$, (c) two electron valleys near $A=A_0$, and (d) two electron valleys.}
\end{figure}

However, another possibility exists. The {\it simple} semiclassical theory fails to explain the anomalous LMR and PHE because most semiclassical formulae assume a spherical or an ellipsoidal Fermi surface and do not consider the fine structure of the band dispersion of carriers \cite{Jones-Zener,Seitz1950,MacKey1969}. The detailed evaluation of the characteristic band structure may lead to the correct solution. The formula derived by Chambers accounts for the arbitrary shape of the Fermi surface in terms of the velocity, i.e., up to the first-order derivative of energy dispersion for $\bm{k}$ \cite{Chambers,Quantum_Kittel}. Recently, the semiclassical formula provided by Mackey-Sybert was extended to account for the arbitrary shape of the Fermi surface through the $\bm{k}$-dependent effective-mass, i.e., up to the second-order derivative of energy dispersion (Eq. \eqref{eq_aws}) \cite{Awashima2019}. The extended Mackey-Sybert formula can explain the LMR even with a single closed Fermi surface \cite{Awashima2019}, which is experimentally well-known but has never been theoretically obtained based on the {\it simple} semiclassical theory. The key to explain the LMR with a single closed Fermi surface lies in the off-diagonal effective-mass, which reflects the detailed geometrical characteristics of the Fermi surface. The $\bm{k}$-dependent effective-mass can yield galvanomagnetic effects that cannot be predicted based on the {\it simple} semiclassical theory \cite{Awashima2019}.

Herein, we show that the negative LMR and the sign change in the PHE are realized in Weyl semimetals based on the extended Mackey-Sybert formula, even without topological effects, the Berry curvature, or the chiral anomaly. Among these anomalous galvanomagnetic effects, the off-diagonal effective-mass, which is singular in Weyl semimetals, plays a crucial role.

We employ the standard model of Weyl semimetals, which is expressed using the following Hamiltonian \cite{Lu2015,Wang2016}:
\begin{equation}
    H = A\left(k_x\sigma_x+k_y\sigma_y\right) + M\left(k_w^2-k^2\right)\sigma_z,
    \label{Hamiltonian}
\end{equation}
where $\sigma_{x,y,z}$ are the Pauli matrices, $M$ and $k_w$ are the model parameters, and $k^2 = k_x^2+k_y^2+k_z^2$. $A$ represents the strength of the Weyl nature, which couples the electron and hole bands. $M$ denotes the inverse mass of the electron and hole bands.
The eigenenergy of this Hamiltonian is 
$
E(\bm{k}) =\pm \sqrt{M^2(k_w^2-k^2)^2+A^2(k_x^2+k_y^2)}.
$
Eq. \eqref{Hamiltonian} is termed the ``Weyl semimetal’’ model because the first term corresponds to two-dimensional Weyl electrons and the second term corresponds to the ordinary semimetals with free electrons and holes, the extrema of which are located at the $\Gamma$-point ($\bm{k}=0$). 
The energy dispersions for $M=16$ eV \AA$^2$ and $k_w= 0.01$ \AA$^{-1}$ are shown in the left panel of Fig. \ref{fig1}. The bands cross linearly at the Weyl point, $\bm{k} = (0, 0, \pm k_w)$. The gap opens at $\bm{k}=(\pm k_w, \pm k_w, 0)$ with increasing $A$. The charge neutrality is maintained only when the Fermi energy $E_F$ is located at the Weyl point. In practice, $E_F$ rarely coincides with the Weyl point. Therefore, hereinafter, we assume $E_F$ to be located slightly above the Weyl point, $E_F=1$ meV, where the electron carriers are slightly larger than the hole carriers. The conclusions are independent of the sign of $E_F$ in our model.
The Fermi surfaces are shown in Fig. \ref{fig1} for $A/{\rm eV\AA}=0.02$, 0.105, 0.11, and 0.24.
For small $A$ values, a large electron and a small hole spherical Fermi surface appear concentrically. In other words, the system is semimetallic. As $A$ increases, the electron Fermi surface becomes dented around the equator, whereas the hole Fermi surface swells. The electron surface touches the hole surface along the equator at $A_0=[2M^2 k_w^2 -2\sqrt{M^2(M^2k_w^4 -E_F^2)}]^{1/2}$ ($\simeq 0.106$ eV\AA\ in the present case). For $A>A_0$, the Fermi surface is separated into two electron pockets. Specifically, the topology of the Fermi surface changes (the Lifshitz transition) from a semimetal ($A<A_0$) to two electron valleys ($A>A_0$).

First, we estimate the MR and PHE using the classical formula of magneto-conductivity \cite{Quantum_Kittel,Ziman,Beer1963,SeeSM}.
In the semimetal region ($A<A_0$) [Fig. \ref{fig1} (a)], the isotropic electron and hole carriers exhibit sizable TMR and no LMR. Accordingly, the $\pi$-period PHE appears because the PHE is generally
$
 \rho_{\rm PHE}=-\Delta \rho_{\rm diff}\sin \theta \cos \theta,
$
where $\Delta \rho_{\rm diff}=\rho_\perp-\rho_\parallel$. 
($\rho_\perp$ and $\rho_\parallel$ are the resistivity under the magnetic field perpendicular and parallel to the electric current, respectively.)
By contrast, in the two-electron-valley region $A>A_0$ [Fig. \ref{fig1} (d)], the Fermi surfaces can be approximated to two parallel ellipsoids, which are equivalent to a double-size single ellipsoid. In such a case, both TMR and LMR (and so PHE) will vanish.
Thus, it is naïvely expected that the amplitude of TMR and PHE would decrease as $A$ increases and that LMR would remain vanishingly small.

Now, we accurately calculate the magneto-conductivity tensor based on the semiclassical Boltzmann theory. The semiclassical formula derived by Mackey and Sybert \cite{MacKey1969} has been extended to consider the arbitrary shape of the Fermi surface in the following form \cite{Awashima2019}:
\begin{eqnarray}\label{eq_aws}
 \sigma_{\lambda,\mu} &=& e\left\langle v_{\lambda} 
 \left\{{\bm v}\cdot
 [(e\tau)^{-1} - {\hat B}\cdot{\hat \alpha_{\bm k}}]^{-1} \right\}_\mu \right\rangle_F, 
\end{eqnarray}
where $e$ ($>0$) is the elementary charge, $\tau$ is the relaxation time, ${\bm v}_{\bm k} = \nabla_{\bm k} E/\hbar$ is the velocity, and $\alpha_{{\bm k}\mu\nu} = \hbar^{-2}\partial^2 E/ \partial k_\mu \partial k_\nu$ is the inverse of the $\bm{k}$-dependent effective-mass tensor, which represents the curvature of the equi-energy surface. $\hat{B}$ is the magnetic field tensor given as $B_{\lambda\mu} = -\epsilon_{\lambda \mu \nu} B_\nu$ ($\epsilon_{\lambda \mu \nu}$: the Levi-Civita symbol) \cite{MacKey1969}.
$\langle \cdots \rangle_F = \int ({\bm d}{\bm k}/4\pi^3) \cdots (-\partial f_0/\partial E)$ corresponds to integration along the Fermi surface at low temperatures, where $f_0$ is the Fermi distribution function without the magnetic field.

Based on the extended Mackey-Sybert formula, Eq. \eqref{eq_aws}, the coefficient of the magnetic field is generally an effective-mass tensor, i.e., the electron's orbital motion is coupled with the magnetic field through the effective-mass.
Importantly, this formula does not include any topological effects, such as the Berry curvature and the chiral anomaly. The unusual behaviors of galvanomagnetic effects, as discussed subsequently, are simply due to the nonuniform effective-mass in Weyl semimetals.

\begin{figure}[htb]
\includegraphics[width=6cm]{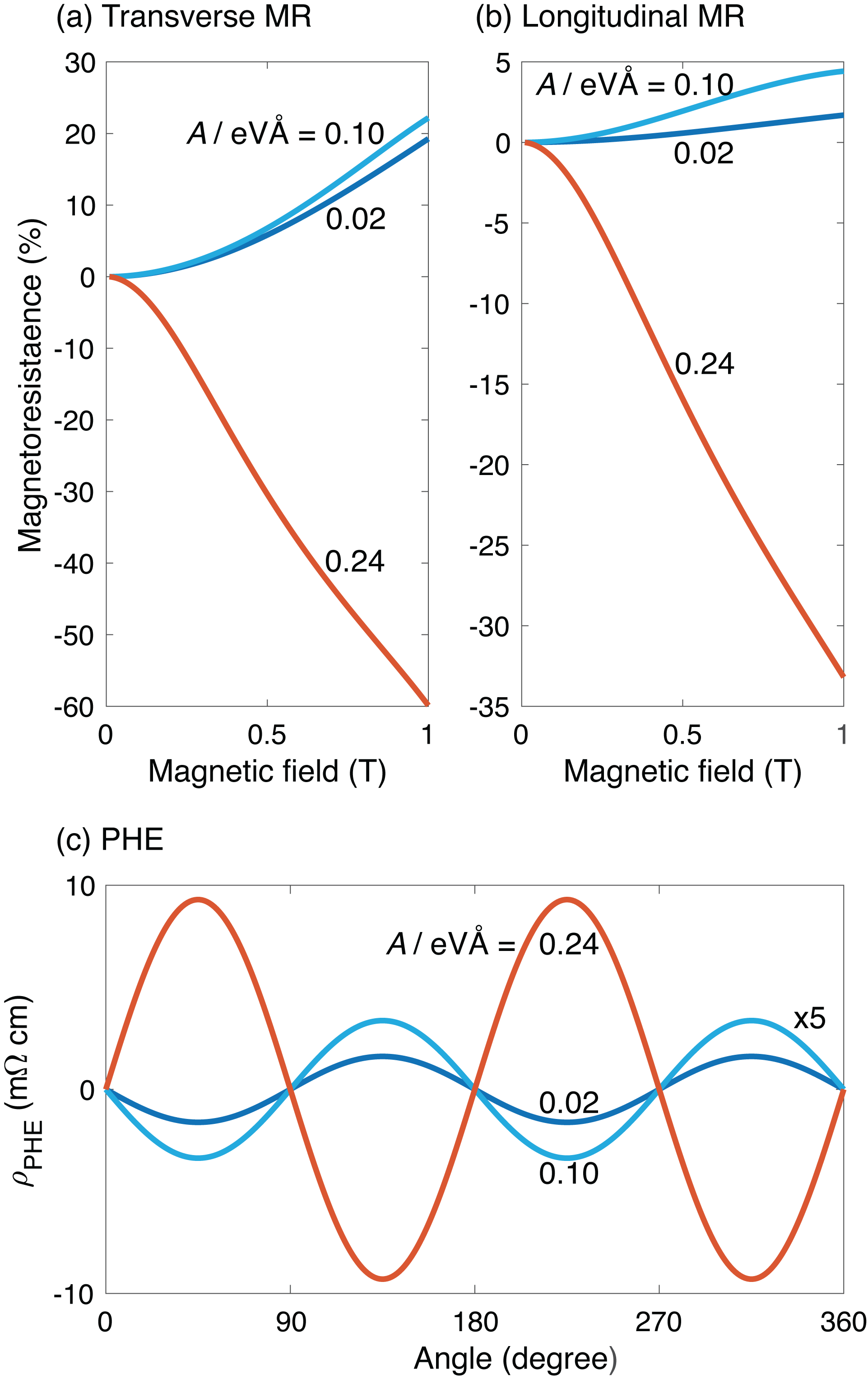}
\caption{\label{fig2} Magnetic field dependence of (a) transverse and (b) longitudinal MR for $A/{\rm eV\AA}=0.02, 0.10, 0.24$. (c) Angular dependence of the PHE at $B=0.5$ T. The PHEs for $A/{\rm eV\AA}=0.02, 0.10$ are amplified fivefold. The temperature is fixed at $T=0.2$ K for (a)-(b).}
\end{figure}

\begin{figure}[t]
\includegraphics[width=7cm]{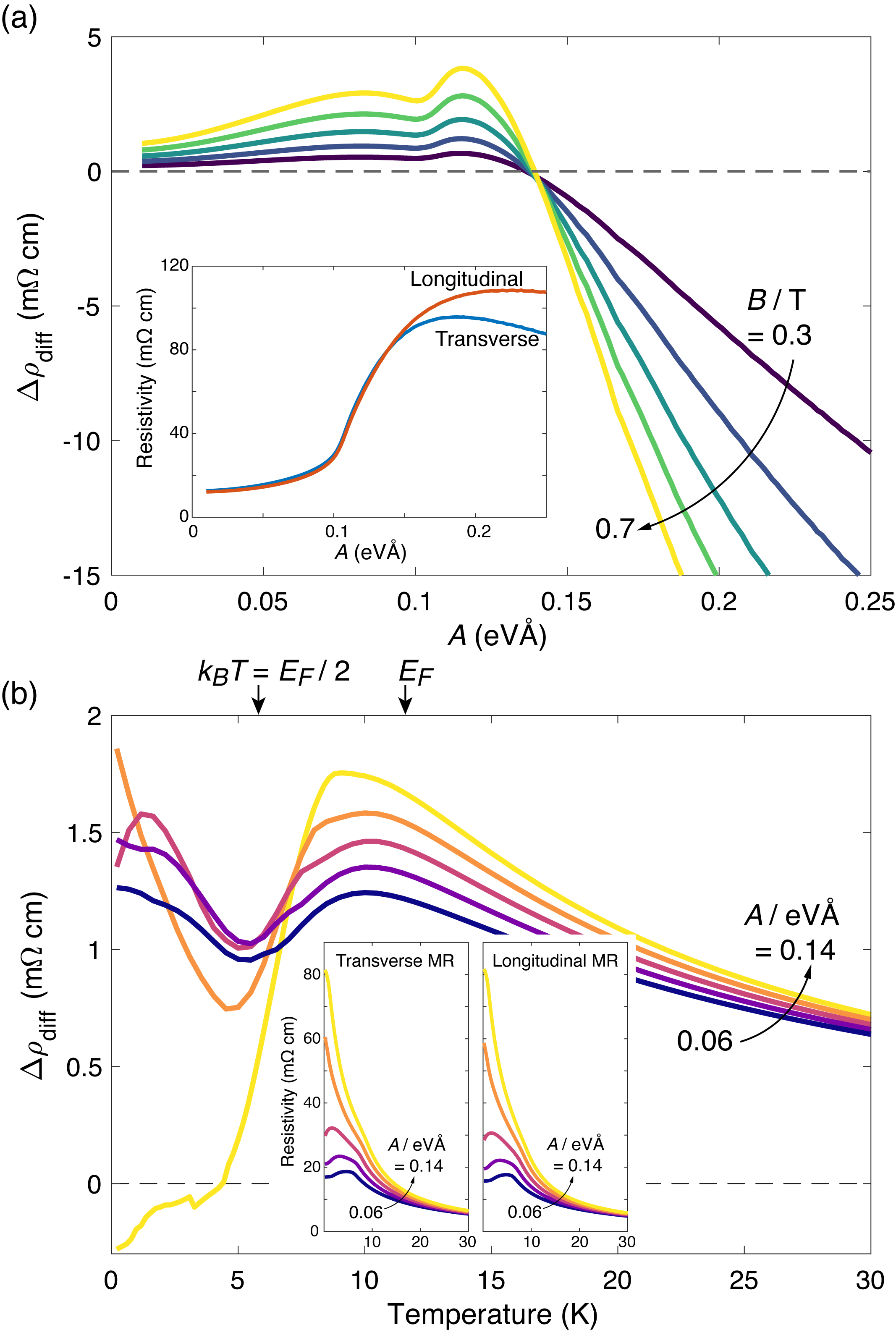}
\caption{\label{fig3} (a) Difference between TMR and LMR, $\Delta \rho_{\rm diff}=\rho_\perp - \rho_\parallel$ as a function of $A$ at $T=0.2$ K for $B=0.3$-0.7 T. The inset shows the $A$-dependence of $\rho_{\perp, \parallel}$ at 0.5 T.
(b) Temperature dependence of $\Delta \rho_{\rm diff}$ at $B=0.5$ T for $A=0.06$-0.14 eV\AA. The inset shows the $T$-dependence of $\rho_{\perp, \parallel}$.}
\end{figure}

The results obtained using Eq. \eqref{eq_aws} are shown in Fig. \ref{fig2} (In Figs. \ref{fig2} and \ref{fig3}, we show the results with $\tau=1.0$ ps). 
In the present study, the current orientation was fixed along the $x$-direction, and the magnetic field was rotated in the $x$-$y$ plane as $\bm{B}=(B\cos\theta,B\sin\theta,0)$. We thus obtained the LMR and TMR for $\theta=0$ and $\pi/2$.
In the semimetallic region ($A=0.02$ eV\AA), a sizable TMR is achieved. Its field dependence is $\Delta \rho_\perp \propto B^2$, which is a typical MR property in semimetals. Near the Lifshitz transition ($A=0.10$ eV\AA), $\Delta \rho_\perp$ is almost unchanged, although the saturation field is rather low. Thus far, the properties of $\Delta \rho_\perp$ are consistent with our classical estimation.
In the two-valley region, however, $\Delta \rho_\perp$ exhibits a significant negative MR, contradicting the classical estimation.
The behavior of LMR is almost the same as that of TMR, i.e., $\Delta \rho_\parallel \propto B^2$ in the semimetallic region and near the Lifshitz transition, whereas $\Delta \rho_\parallel <0 $ in the two-valley region. (Note that $ |\Delta \rho_\parallel |< |\Delta \rho_\perp| $.) The observations regarding the LMR contradict the classical theory, which predicts $\Delta \rho_\parallel = 0$ for the entire range of $A$.
The PHE exhibits an angular dependence of $-\sin 2\theta$ in the semimetallic region ($A\lesssim A_0$), whereas the sign of the angular oscillation is inverted in the two-valley region ($A\gtrsim A_0$). The origin of this sign change can be easily explained by plotting $\Delta \rho_{\rm diff}$, as shown in Fig. \ref{fig3} (a). The reduction in $\rho_\perp$ is larger than $\rho_\parallel$; hence, $\rho_\perp $ becomes smaller than $\rho_\parallel$ near $A=0.14$ eV\AA, resulting in the sign change in $\Delta \rho_{\rm diff}$.

The negative LMR and TMR in the two-valley region is rather surprising. The simple semiclassical theory does not predict a negative MR even if anisotropy or multicarrier effects are considered \cite{Beer1963,Zhu2018,Mitani_2020}. This negative LMR and TMR in the two-valley region can be attributed to the off-diagonal effective-mass, which have not been considered in the simple semiclassical theory. The best approach for understanding the significance of the off-diagonal effective-mass is to find the Jones-Zener expansion of the extended Mackey-Sybert formula, Eq. \eqref{eq_aws} \cite{Fuchser_1970}. By expanding $[1/e\tau -\hat{B}\cdot \hat{\alpha}_{\bm k}]^{-1}$ in terms of $B$, the second-order term of LMR for $\bm{B}=(B,0,0)$ can be expressed as 
\begin{align}
    \rho_{xx}^{(2)}=\tau \Lambda_x^2
    &\langle
    v_z v_x\left(\alpha_{zx}\alpha_{yy}-\alpha_{xy}\alpha_{yz}\right)
    \nonumber\\&
    + v_x v_y \left( \alpha_{xy}\alpha_{zz} -\alpha_{yz}\alpha_{zx} \right)
    \rangle_F B^2,
    \label{LMR2nd}
\end{align}
where $\Lambda_\mu = 1/\langle v_\mu^2 \rangle_F$.
The first term in the parentheses, which include one off-diagonal and one diagonal effective-mass, $\alpha_{\lambda \mu}\alpha_{\nu \nu}$, dominates the second term, which includes two off-diagonal effective-masses. This is because the diagonal term is usually larger than the off-diagonal one. Eq. \eqref{LMR2nd} shows that negative LMR can appear when the off-diagonal effective-mass becomes negative. Such a situation does occur in Weyl semimetals. 

\begin{figure}[tb]
\includegraphics[width=9cm]{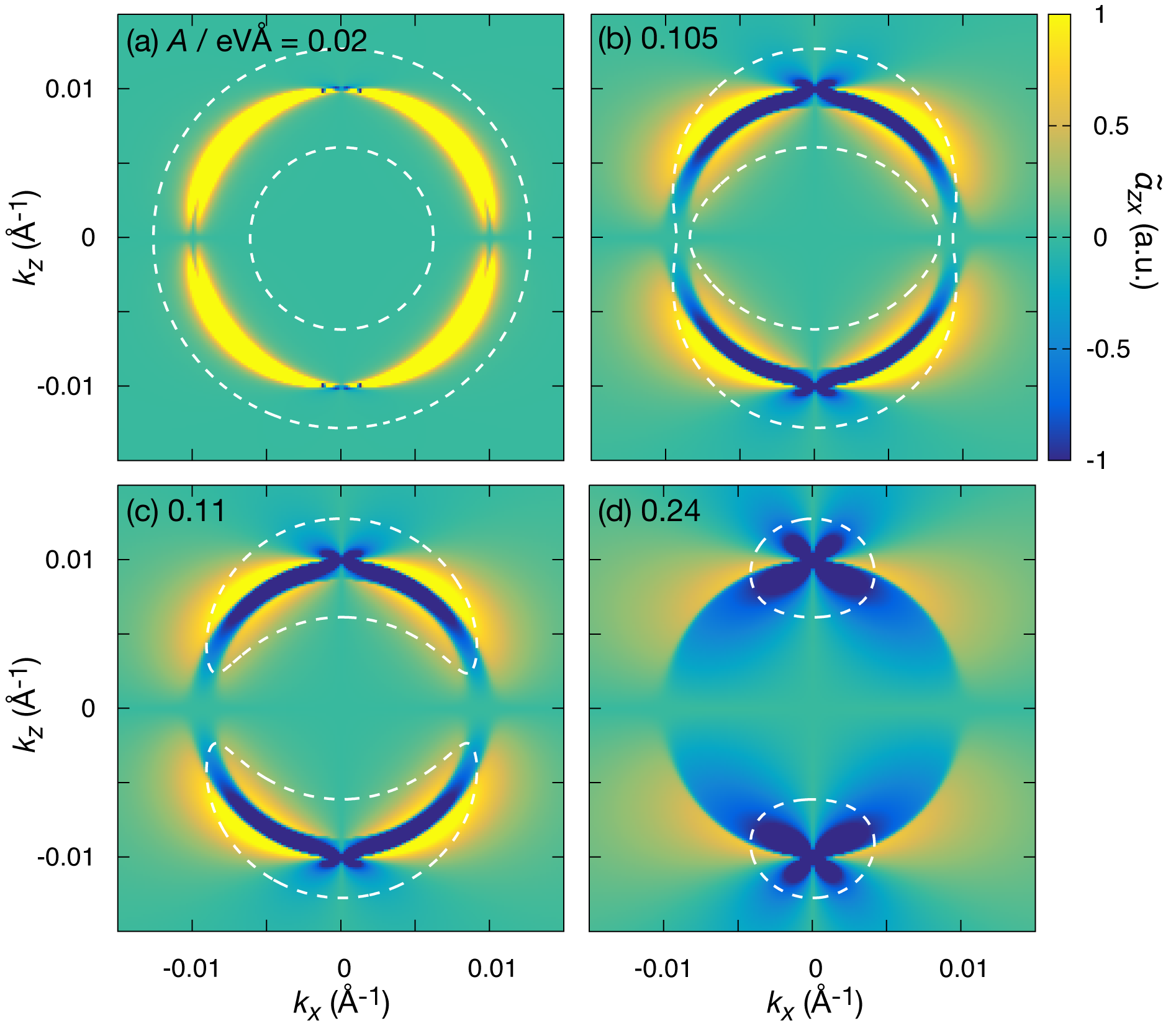}
\caption{\label{fig4} Off-diagonal effective-mass $\tilde{\alpha}_{zx}(k_x, 0, k_z)$ for $A/{\rm eV\AA}=0.02, 0.105, 0.11, 0.24$. The blue regions correspond to $\tilde{\alpha}_{zx}<0$. The dashed white lines indicate the Fermi surface. The Weyl points are located at $(k_x, k_z)=(0, \pm 0.01 {\rm \AA}^{-1})$.}
\end{figure}

The off-diagonal effective-mass $\alpha_{zx}$ in the Weyl semimetal model is plotted for $A/{\rm eV\AA}=0.02$, 0.105, 0.11, and 0.24 in Fig. \ref{fig4}. Although $\alpha_{zx}$ is an odd function with respect to $k_z$ and $k_x$, it appears in Eq. \eqref{LMR2nd} as $v_z v_x \alpha_{zx}$, which is an even function. Thus, in Fig. \ref{fig4}, we plot $\tilde{\alpha}_{zx}=\alpha_{zx} v_z v_x /|v_z||v_x|$, which reveals the unusual sign change of the off-diagonal effective-mass. The dashed lines indicate the corresponding Fermi surfaces. The blue region corresponds to negative $\tilde{\alpha}_{zx}$. For the semimetallic region (a), the negative-$\tilde{\alpha}_{zx}$ portion is narrow, and the $\tilde{\alpha}_{zx}$ on the Fermi surface is positive for the entire region. The negative-$\tilde{\alpha}_{zx}$ region grows as $A$ increases. Around the Lifshitz transition (b), (c), a part of the Fermi surface has the negative $\tilde{\alpha}_{zx}$, but the positive region remains dominant. In the two-valley region (d), almost the entire Fermi surface has negative $\tilde{\alpha}_{zx}$, which is the dominant contributor to the MR. These results explain the negative LMR in the two-valley region.

%The Gaussian curvature is known to appear in various physical quantities, such as diamagnetism \cite{Peierls1933,Hebborn1959} and the spin Hall effect \cite{Fuseya2015}. In such phenomena where the orbital motion of electrons couples with the magnetic field or the magnetic moment, their macroscopic response coefficients are given by the Gaussian curvature. MR (and the subsequent PHE) is one such magneto-orbital phenomenon.
In the case of the TMR, $\bm{B}=(0,B,0)$, the second-order term is
\begin{align}
    \rho_{xx}^{(2)}&= \Bigl[ \tau \Lambda_x^2 \langle
    v_x^2 \left(\alpha_{xx}\alpha_{zz}-\alpha_{zx}^2 \right)
    \rangle_F 
\nonumber\\&
    -\tau \Lambda_x^2 \Lambda_z \left\langle
    v_x v_z \alpha_{zx}-v_x^2\alpha_{zz}
    \right\rangle_F^2
    \Bigr] B^2,
\end{align}
where the first term originates from the diagonal conductivity, $\sigma_{xx}$, and the second one originates from the off-diagonal Hall conductivity, $\sigma_{zx}$. 
%Note that $\alpha_{xx}\alpha_{zz}-\alpha_{zx}^2$ corresponds to the Gaussian curvature of the Fermi surface. 
The large negative $\alpha_{zx}$ increases $\sigma_{zx}$ and reduces $\sigma_{xx}$. When $\sigma_{zx}$ dominates $\sigma_{xx}$, the negative TMR emerges. Such a condition can actually be satisfied in the case of Weyl semimetals, as shown in Fig. \ref{fig4}.

%Eqs. \eqref{LMR2nd} and \eqref{TMR2nd} are general formulae independent from the model. Therefore, our conclusion that the negative off-diagonal effective-mass yields the negative LMR is an entirely general statement.

Some previous studies suggest that negative LMR is evidence of the chiral anomaly \cite{Xiaochun2015,Li2016,Liang2018,Vashist2021}.
In contrast, the present calculation shows that the negative off-diagonal effective-mass can cause a negative LMR, which is inherent in Weyl semimetals. In the present model, where the effective-mass is isotropic in the $x$-$y$ plane, the negative LMR is accompanied by the negative TMR. This is due to the oversimplification of the model of Weyl semimetals. $\Delta \rho_\perp >0$ and $\Delta \rho_\parallel <0$ can be obtained by considering the detailed band structure of Weyl semimetals. For example, $\Delta \rho_\perp >0$ and $\Delta \rho_\parallel <0$ were obtained for the model of SrTiO$_3$ using the extended Mackey-Sybert formula \cite{Awashima2019}.

Finally, we discuss the temperature $T$ dependence of the MR.
The $T$-dependence of $\Delta \rho_{\rm diff}$ is depicted in Fig. \ref{fig3} (b). To observe the effects of the Weyl dispersion as transparently as possible, we assumed $\tau$ to be constant for $T$ (note that the anomalous $T$-dependence shown below remains even if we consider the $T$-dependence of $\tau$, which should be a monotonic function of $T$). 
$\Delta \rho_{\rm diff}$ is generally expected to be a monotonically decreasing function.
However, as shown in Fig. \ref{fig3} (b), $\Delta \rho_{\rm diff}$ exhibits a dip near $T=5$ K.
The origin of this dip also lies in the negative off-diagonal effective-mass.
As shown in Fig. \ref{fig4}, the off-diagonal effective-mass is singular and changes its sign around the Weyl points. Therefore, the region around the Weyl point hinders the MR when the tail of the Fermi distribution function reaches the Weyl point; this produces the aforementioned dip.
As $A$ increases, the dip becomes deeper because the impact of the negative off-diagonal effective-mass is amplified. For a sufficiently large $A$, the dip becomes sufficiently deep for $\Delta \rho_{\rm diff}$ to be negative, even at the zero-temperature limit.
The temperature at which $\Delta \rho_{\rm diff}$ attains its maximum value approximately corresponds to the energy difference between the Weyl point and $E_F$.
The dip appears at around $k_B T=E_F/2$; it can be seen more clearly in $\Delta \rho_{\rm diff}$ than in TMR or LMR, where the anomaly is masked by the background $T$-dependence, as shown in the inset of Fig. \ref{fig3} (b).
Based on this distinctive property, the energy scale of the Weyl point can be determined by measuring the $T$-dependence of $\Delta \rho_{\rm diff}$. 
%Both of the LMR and TMR is inversely proportional to the carrier number, hence the decay of $\Delta \rho$ at higher than 10 K is simply due to increasing of number in thermally excited carrier.

In summary, we investigated galvanomagnetic effects on the basis of the Boltzmann theory, focusing on the geometrical characteristics of the Fermi surface. We identified that the negative off-diagonal effective-mass, which is inherent in Weyl semimetals, causes the negative MR and a sign change in the PHE. 
Our results highlight the critical role the off-diagonal effective-mass plays in the galvanomagnetic effects. The results of this study will provide significant assistance in obtaining a more intuitive understanding of the anomalous galvanomagnetic effect. For example, one can realize anomalies in the galvanomagnetic effect without calculating the resistivity but just by calculating the effective-mass, which is easily obtained from the band calculations.
Another direct consequence of the negative off-diagonal effective-mass is the anomalous dip in the $T$-dependence of $\Delta \rho_{\rm diff}$ or the amplitude of the PHE.
By measuring the dip structure, we can evaluate the energy difference between the Weyl point and the Fermi energy, which is crucial for the study of Weyl semimetals. %; however, this aspect is frequently overlooked.
Non-monotonic $T$-dependence (including the dip and sign change) has been observed in actual materials \cite{Kumar2018,Yang2020,Wu2018,Li2018}. This dependence can originate from the negative off-diagonal effective-mass. To clarify, the effective-mass (and subsequently the galvanomagnetic effect) should be calculated based on the detailed band-structure. Future research should, therefore, focus on more detailed investigations into this.

We thank B. Fauqu\'e and Z. Zhu for helpful comments.
This work is supported by JSPS KAKENHI (Grant No. 19H01850).

\bibliography{PHE}
\end{document}